\documentclass[twocolumn,showpacs,prb]{revtex4}

\usepackage{amssymb}
\usepackage{graphicx}
\usepackage{hyperref}

\def\gz{\ifmmode{Z\hskip -4.8pt Z}
    \else{\hbox{$Z\hskip -4.8pt Z$}}\fi}

\newcommand{\be}{\begin{equation}}
\newcommand{\ee}{\end{equation}}
\newcommand{\bea}{\begin{eqnarray}}
\newcommand{\eea}{\end{eqnarray}}

\begin{document}

\title{Effective Hamiltonian for transition-metal compounds. Application to
Na$_x$CoO$_2$}
\author{A.~A.~Aligia}
\email{aligia@cab.cnea.gov.ar}
\affiliation{Centro At\'omico Bariloche and Instituto Balseiro, Comisi\'on Nacional de
Energ\'{\i}a At\'omica, 8400 Bariloche, Argentina}
\author{T.~Kroll}
\affiliation{IFW Dresden, P.O. Box 270016, D-01171 Dresden, Germany}
\date{\today}

\begin{abstract}
We describe a simple scheme to construct a low-energy effective Hamiltonian $%
H_{\mathrm{eff}}$ for highly correlated systems containing non-metals like
O, P or As (O in what follows) and a transition-metal ($M$) as the active
part in the electronic structure, eliminating the O degrees of freedom from
a starting Hamiltonian that contains all $M$ d orbitals and all non-metal p
orbitals. We calculate all interaction terms between d electrons originating
from Coulomb repulsion, as a function of three parameters ($F_0$, $F_2$ and $%
F_4$) and write them in a basis of orbitals appropriate for cubic,
tetragonal, tetrahedral or hexagonal symmetry around $M$. The approach is
based on solving exactly (numerically if necessary) a $M$O$_n$ cluster
containing the transition-metal atom and its $n$ nearest O atoms (for
example a CoO$_6$ cluster in the case of the cobaltates, or a CuO$_n$
cluster in the case of the cuprates, in which $n$ depends on the number of
apical O atoms), and mapping them into many-body states of the same symmetry
containing d holes only. We illustrate the procedure for the case of Na$_x$%
CoO$_2$. The resulting $H_{\mathrm{eff}}$, including a trigonal distortion $D
$, has been studied recently and its electronic structure agrees well with
angle-resolved photoemission spectra [A. Bourgeois, A. A. Aligia, and M. J.
Rozenberg, Phys. Rev. Lett. \textbf{102}, 066402 (2009)]. Although $H_{%
\mathrm{eff}}$ contains only 3d $t_{2g}$ holes, the highly correlated states
that they represent contain an important amount not only of O 2p holes but
also of 3d $e_{g}$ holes. When more holes are added, a significant
redistribution of charge takes place. As a consequence of these facts, the
resulting values of the effective interactions between $t_{2g}$ states are
smaller than previously assumed, rendering more important the effect of $D$
in obtaining only one sheet around the center of the Brillouin zone for the
Fermi surface (without additional pockets).
\end{abstract}

\maketitle

\section{Introduction}

In condensed matter physics, several materials containing transition metal
atoms ($M$) and oxygen have attracted great interest, for example the
superconducting cuprates, manganites with high magnetoresistance, and more
recently cobaltates and Fe pnictides. In the latter, the role of O in the
electronic structure is replaced by either P or As. The electronic structure
of these systems near the Fermi energy is determined essentially by some d
orbitals of $M$ and some 2p orbitals of O (or 3p, 4p for P, As in the
pnictides), while the rest of the orbitals and elements play the role of
reservoirs of charge. In most of these systems, strong correlations
at $M$ play an essential role. For example, while first-principles
calculations predict that the undoped cuprates like La$_{2}$CuO$_{4}$ are
non-magnetic metals,\cite{matt} they are in fact antiferromagnetic
insulators. In the cobaltates Na$_{x}$CoO$_{2}$, first-principles
calculations \cite{lda} predicted a Fermi surface with six prominent hole
pockets along the $\Gamma -K$ direction, which are absent in measured
angle-resolved photoemission spectra.\cite{arpes1,arpes2}

Because of the presence of strong correlations, numerical methods which
treat exactly the interactions in a small cluster of the system have led to
a considerable advance in the understanding of these materials.\cite%
{dago,pair,hored,zem,ralko} Due to the exponential increase of the size of
the Hilbert space, it is highly desirable that the set of relevant orbitals
included in these numerical studies is the smallest possible. For the
cuprates, low-energy reduction procedures that eliminate the O degrees of
freedom, simplifying the problem to an effective one-band one,\cite%
{jef,schu,brin,bel,sys,hub,opti} have been very successful, in spite of the
fact that doped holes enter mainly at O atoms \cite{nuck,kuip,pelle}
(optical properties related with O atoms were calculated using these
one-band models \cite{opti,erol}). Approximate treatments of the problem are
also expected to lead to more accurate results when applied to the effective
model, because the construction of the latter begins with an exact treatment of
the local interactions at $M$. For example a slave-boson mean-field
treatment applied to the three-band model (with Cu and O orbitals) for the
cuprates predicts a metal for the most accepted parameters,\cite{bal}
whereas the same treatment on the one-band effective model gives a gap that
agrees with experiment.\cite{brin}

The main idea behind the construction of an effective one-band model for the
cuprates, proposed by Zhang and Rice,\cite{zr} starts from the exact
solution of a CuO$_{4}$ cluster with one and two holes added to the vacuum
in which the Cu atoms are in the 3d$^{10}$ configuration and the O atoms in
the 2p$^{6}$ one. For two holes, the ground state is a singlet formed
essentially by one hole in the 3d$_{x^{2}-y^{2}}$ orbital of Cu and one hole
in the linear combination of 2p orbitals of the nearest-neighbor O sites,
with the same symmetry. Mapping this singlet state onto the corresponding
state of a one-band model leads to the usual Hubbard or $t-J$ models. Using
a mapping involving non-orthogonal singlets, Zhang has shown that in a
particular case, the mapping is exact.\cite{zhang} More systematic
derivations lead to additional terms of smaller magnitude,\cite{sys,hub,vali}
like three-site terms, which might be important for superconductivity.\cite%
{rvb,manu,pair,suphu} Three-site terms play also an important role in
effective models for vanadates or other systems with $t_{2g}$ electrons.\cite%
{dagh} Triplet states can be included perturbatively.\cite{trip}

Second-order perturbation theory in the Cu-O hopping applied to the
three-band model for the cuprates leads to an interaction between Cu spins
and O holes.\cite{spife} Adding to it the superexchange interaction between
Cu spins one has the so called spin-fermion model. Similar derivations have
been made for NiO,\cite{oles} doped Y$_{2}$BaNiO$_{5}$,\cite{epl,com} and
other perovskites.\cite{feiner} An important point is that the parameters of
the model are improved if instead of taking the values that result from
perturbation theory, they are obtained fitting the energy levels of a CuO$%
_{4}$ cluster in the case of the cuprates \cite{spife} or a NiO$_{6}$
cluster in the case of the nickelates.\cite{epl} The Cu and O photoemission
spectra for the cuprates obtained from the resulting spin-fermion model are
practically identical to those of the three-band Hubbard model but required
a much less computational effort for the same cluster.\cite{spife}

The above mentioned results suggest that in general the key for constructing
a low-energy Hamiltonian is to solve exactly a cluster $M$O$_{n}$ containing
the transition metal atom and the O atoms of its neighborhood. In fact, the
cell-perturbation method \cite{jef} used extensively in the cuprates \cite%
{jef,brin,bel,bel2,hub,opti} exploits this idea: it divides the
system into orthogonal cells containing one $M$ atom each, using
appropriate O Wannier functions centered at $M$, solves exactly
each cell, writes the Hamiltonian in the basis of the resulting
eigenstates, retains only the relevant low-energy eigenstates, and
includes the rest perturbatively. However, while in the cuprates
the O Wannier functions can be easily constructed and the Hilbert
space of the cluster involves only a few eigenstates, in the
general case, as we show below, at least ten d and p orbitals are
involved. In this case, the size of the Hilbert space of the
$M$O$_{n}$ cluster is $4^{10}$ and (except for favorable cases)
the matrices should be solved numerically. For O rich systems, we
propose to neglect the overlap between different O clusters and
retain only the most important terms in the resulting effective
model $H_{\mathrm{eff}}$ (the effect of this overlap in the
cuprates was calculated in Ref. \onlinecite{sys}). A discussion of
the effect of the overlap in the cobaltates is contained in
Section \ref{overlap}. Similarly to the above mentioned ideas, the
most relevant parameters are obtained from a fit of the low-energy
levels of the original $M-$O \ multiband model $H$ (obtained
numerically) and those of $H_{\mathrm{eff}}$ in the $M$O$_{n}$
cluster. An additional calculation is required to obtain effective
hoppings between clusters.

An essential ingredient of the original multiband model $H$ is the part of
it which contains the interactions between 3d electrons $H_{I}$. While it is
straightforward to obtain it %assuming spherical symmetry of $H_{I}$
using known results of atomic physics, the derivation is lengthy and it
seems that only simplified forms were used so far for research in these
systems. For example in a recent study of Fe pnictides,\cite{kane} a
simplified expression derived previously \cite{oles2} was used. While the
form of $H_{I}$ is well known when either only $e_{g}$ orbitals \cite{epl}
or only $t_{2g}$ orbitals \cite{fre,gus} are important, the correct
expressions were not always used.\cite{com,fre} In the general case, new
terms appear which were not discussed before. One of the goals of this work
is to present the complete $H_{I}$. We believe that this will be useful for
future theoretical work on strongly correlated systems of transition metals,
when the relevant orbitals cannot be restricted to either $e_{g}$ or $t_{2g}$
only.

In this work, we outline the derivation of $H_{\mathrm{eff}}$ as described
above. While for the sake of clarity we consider that the relevant non-metal
orbitals are the 2p of O atoms, they could also be the 3p of P or 4p of As
in the case of the Fe pnictides. We consider in detail the specific example
of the cobaltates Na$_{x}$CoO$_{2}$. A brief description of this case,
together with a dynamical-mean field treatment of $H_{\mathrm{eff}}$ to
study the electronic structure and Fermi surface of the system was published
before.\cite{dmft}

In Section II, we describe the construction of the multiband model
containing both $M$ 3d (or 4d) and O 2p (or P 3p, As 4p) electrons,
including all the d-d interaction terms. In Section III, we derive the
effective Hamiltonian $H_{\mathrm{eff}}$, which contains only effective d
operators but no O ones. While some considerations are valid for the general
case, the explicit construction is done for the case of Na$_{x}$CoO$_{2}$.
Section IV is a summary and discussion.

\section{The starting Hamiltonian}

As in many transition-metal compounds, we assume that the essential part of
the electronic structure consists of the d electrons of the transition-metal
atoms and the p electrons of the O (or other non-metal) atoms. An example in
which only one d orbital is relevant is the three-band Hubbard model for the
cuprates,\cite{eme,var} with parameters determined by
constrained-density-functional theory.\cite{cdf1,cdf2} The Hamiltonian
contains d-p hopping terms and interactions. The most important of the
latter are the interaction terms among the d electrons.

\subsection{The interactions inside the d shell}

Here we can consider only one transition-metal atom and drop the site index
for simplicity. The part of the Hamiltonian that contains the interaction
among the 10 d spin-orbitals is \cite{nege}

\begin{equation}
H_{I}=\frac{1}{2}\sum_{\lambda \mu \nu \rho }V_{\lambda \mu \nu \rho
}d_{\lambda }^{+}d_{\mu }^{+}d_{\rho }d_{\nu },  \label{hi0}
\end{equation}%
where $d_{\lambda }^{+}$ creates an electron or a hole at \ the spin-orbital
$\lambda $ ($H_{I}$ is invariant under an electron-hole transformation) and

\begin{equation}
V_{\lambda \mu \nu \rho }=\int d\mathbf{r}_{1}d\mathbf{r}_{2}\bar{\varphi}%
_{\lambda }(\mathbf{r}_{1})\bar{\varphi}_{\mu }(\mathbf{r}_{2})\frac{e^{2}}{|%
\mathbf{r}_{1}-\mathbf{r}_{2}|}\varphi _{\nu }(\mathbf{r}_{1})\varphi _{\rho
}(\mathbf{r}_{2}),  \label{integ}
\end{equation}%
where $\varphi _{\lambda }(\mathbf{r}_{1})$ is the wave function of the
spin-orbital $\lambda $.

In the basis of given angular momentum and spin projections ($\lambda \equiv
m^{\lambda },\sigma ^{\lambda }$), the orbital part of the wave functions
can be written as $R(r)Y_{2}^{m}(\theta ,\phi )$, where $R$ is the radial
part and $Y_{l}^{m}$ is a normalized spherical harmonic. Using standard
methods of atomic physics \cite{fre,cond} one obtains

\begin{eqnarray}
V_{\lambda \mu \nu \rho } &=&\delta (\sigma ^{\lambda },\sigma ^{\nu
})\delta (\sigma ^{\mu },\sigma ^{\rho })\delta (m^{\lambda }+m^{\mu
},m^{\nu }+m^{\rho })  \nonumber \\
&&\times \sum_{k=0}^{\infty }c^{k}(2m^{\lambda },2m^{\nu })c^{k}(2m^{\rho
},2m^{\mu })\cdot R^{k},  \label{ck}
\end{eqnarray}
with

\begin{eqnarray}
&&c^{k}(lm,l^{\prime }m^{\prime }) =\sqrt{\frac{2}{2k+1}}  \nonumber \\
&&\times \int_{0}^{\pi }P_{k}^{m-m^{\prime }}(\cos \theta )P_{l}^{m}(\cos
\theta )P_{l^{\prime }}^{m^{\prime }}(\cos \theta )\sin \theta d\theta
\nonumber \\
&&R^{k} = e^{2}\int_{0}^{\infty }\int_{0}^{\infty }\frac{r_{<}^{k}}{%
r_{>}^{k+1}}R^{2}(r_{1})R^{2}(r_{2})r_{1}^{2}r_{2}^{2}dr_{1}dr_{2},
\label{r}
\end{eqnarray}%
where $P_{l}^{m}(\cos (\theta ))$ is a normalized Legendre function and $%
r_{<}$ ($r_{>}$) is the smaller (larger) between $r_{1}$ and $r_{2}$. The
values of $c^{k}$ that are needed are tabulated in Ref. \onlinecite{cond}.
To remove uncomfortable denominators, one defines the three free parameters
as $F_{0}=R^{0}$, $F_{2}=R^{2}/49$ and $F_{4}=R^{4}/441$.\cite{fre,cond}

In the presence of a cubic crystal field (point group $O_{h}$) or some other
local symmetry (for example point groups $D_{6h}$, $D_{4h}$, $C_{4v}$, $T_d$), 
it is more convenient to change from the basis of operators with definite
angular momentum projection $d_{m\sigma }^{\dagger}$ to that of irreducible
representations of the point group ($e_{g}$ and $t_{2g}$ for $O_{h}$)
orbitals using

\begin{eqnarray}
d_{\pm 2\sigma }^{\dagger } &=&\frac{1}{\sqrt{2}}(d_{x^{2}-y^{2},\sigma
}^{\dagger }\pm id_{xy,\sigma _{1}}^{\dagger })  \nonumber \\
d_{\pm 1\sigma }^{\dagger} &=& \frac{-1}{\sqrt{2}}(d_{zx,\sigma }^{\dagger
}\pm id_{yz,\sigma }^{\dagger })\text{,}  \nonumber \\
d_{0\sigma }^{\dagger } &=&d_{3z^{2}-r^{2},\sigma }^{\dagger }.
\label{trans}
\end{eqnarray}%
In cubic symmetry, $d_{x^{2}-y^{2},\sigma }^{\dagger}$ and 
$d_{3z^{2}-r^{2},\sigma }^{\dagger }$ correspond to $e_{g}$ symmetry and the
remaining creation operators transform as the $t_{2g}$ irreducible
representation. Each term of the resulting $H_{I}$ contains two creation and
two annihilation operators. $H_{I}$ can be divided in four parts $H_{n}$
according to the number $n$ of $e_{g}$ operators present in each term. There
is no term with only one $t_{2g}$ operator and therefore $H_{3}$ is absent. $%
H_{4}$ is usually enough to describe $e_{g}$ holes in late transition
metals, such as nickelates,\cite{epl} while $H_{0}$ contains the relevant
interaction terms for early transition metals with a few $t_{2g}$ electrons,
such as titanates \cite{fre} or rutenates.\cite{gus} The sums over $\alpha $
($\beta ,\gamma $) below run over the indices of $e_{g}$ ($t_{2g}$)
operators. The interaction can be written as

\begin{equation}
H_{I}=H_{4}+H_{0}+H_{1}+H_{2},  \label{his}
\end{equation}%
with

\begin{eqnarray}
H_{4} &=&U\sum_{\alpha }n_{\alpha ,\uparrow }n_{\alpha ,\downarrow}
\nonumber \\
&+&(U-2J_{e})\sum_{\sigma _{1},\sigma _{2}}n_{x^{2}-y^{2},\sigma
_{1}}n_{3z^{2}-r^{2},\sigma _{2}}  \nonumber \\
&+&J_{e}\sum_{\sigma _{1},\sigma _{2}}d_{x^{2}-y^{2},\sigma _{1}}^{\dagger
}d_{3z^{2}-r^{2},\sigma _{2}}^{\dagger }d_{x^{2}-y^{2},\sigma
_{2}}d_{3z^{2}-r^{2},\sigma _{1}}  \nonumber \\
&+&J_{e}(d_{x^{2}-y^{2},\uparrow }^{\dagger }d_{x^{2}-y^{2},\downarrow
}^{\dagger }d_{3z^{2}-r^{2},\downarrow }d_{3z^{2}-r^{2},\uparrow }  \nonumber
\\
&+&\text{H.c.}),  \label{h4}
\end{eqnarray}

\begin{eqnarray}
H_{0} &=&U\sum_{\beta }n_{\beta ,\uparrow }n_{y\beta ,\downarrow }+\frac{%
U-2J_{t}}{2}\sum_{\beta \neq \gamma }\sum_{\sigma _{1},\sigma _{2}}n_{\beta
,\sigma _{1}}n_{\gamma ,\sigma _{2}}  \nonumber \\
&+&\frac{J_{t}}{2}\sum_{\sigma _{1},\sigma _{2}}\sum_{\beta \neq \gamma
}d_{\beta ,\sigma _{1}}^{\dagger }d_{\gamma ,\sigma _{2}}^{\dagger }d_{\beta
,\sigma _{2}}d_{\gamma ,\sigma _{1}}  \nonumber \\
&+&J_{t}\sum_{\beta \neq \gamma } d_{\beta ,\uparrow }^{\dagger }d_{\beta
,\downarrow }^{\dagger }d_{\gamma ,\downarrow }d_{\gamma ,\uparrow },
\label{h0}
\end{eqnarray}

\begin{eqnarray}
H_{1} &=&\lambda \sum_{\sigma _{1},\sigma _{2}}[\sqrt{3}(d_{x^{2}-y^{2},%
\sigma _{1}}^{\dagger }d_{zx,\sigma _{1}}+\text{H.c.})  \nonumber \\
&\times &(d_{xy,\sigma _{2}}^{\dagger }d_{yz,\sigma _{2}}+\text{H.c.})
\nonumber \\
&-&\sqrt{3}(d_{x^{2}-y^{2},\sigma _{1}}^{\dagger }d_{yz,\sigma _{1}}+\text{%
H.c.})(d_{xy,\sigma _{2}}^{\dagger }d_{zz,\sigma _{2}}+\text{H.c.})
\nonumber \\
&-&2(d_{3z^{2}-r^{2},\sigma _{1}}^{\dagger }d_{xy,\sigma _{1}}+\text{H.c.}%
)(d_{zx,\sigma _{2}}^{\dagger }d_{yz,\sigma _{2}}+\text{H.c.})  \nonumber \\
&+&(d_{3z^{2}-r^{2},\sigma _{1}}^{\dagger }d_{zx,\sigma _{1}}+\text{H.c.}%
)(d_{xy,\sigma _{2}}^{\dagger }d_{yz,\sigma _{2}}+\text{H.c.})  \nonumber \\
&+&(d_{3z^{2}-r^{2},\sigma _{1}}^{\dagger }d_{yz,\sigma _{1}}+\text{H.c.})
\nonumber \\
&\times &(d_{xy,\sigma _{2}}^{\dagger }d_{zx,\sigma _{2}}+\text{H.c.})],
\label{h1}
\end{eqnarray}

\begin{eqnarray}
&&H_{2}=(U-2J_{t})\sum_{\sigma _{1},\sigma _{2}}n_{x^{2}-y^{2},\sigma
_{1}}(n_{zx,\sigma _{2}}+n_{yz,\sigma _{2}})  \nonumber \\
&&+(U-2J_{a})\sum_{\sigma _{1},\sigma _{2}}n_{x^{2}-y^{2},\sigma
_{1}}n_{xy,\sigma _{2}}  \nonumber \\
&&+(U-2J_{b})\sum_{\sigma _{1},\sigma _{2}}n_{3z^{2}-r^{2},\sigma
_{1}}(n_{zx,\sigma _{2}}+n_{yz,\sigma _{2}})  \nonumber \\
&&+(U-2J_{e})\sum_{\sigma _{1},\sigma _{2}}n_{3z^{2}-r^{2},\sigma
_{1}}n_{xy,\sigma _{2}}  \nonumber \\
&&+J_{t}\sum_{\sigma _{1},\sigma _{2}}(d_{x^{2}-y^{2},\sigma _{1}}^{\dagger
}d_{zx,\sigma _{2}}^{\dagger }d_{x^{2}-y^{2},\sigma _{2}}d_{zx,\sigma _{1}}
\nonumber \\
&&\text{ \ \ \ \ \ \ \ \ \ \ \ }+d_{x^{2}-y^{2},\sigma _{1}}^{\dagger
}d_{yz,\sigma _{2}}^{\dagger }d_{x^{2}-y^{2},\sigma _{2}}d_{yz,\sigma _{1}})
\nonumber \\
&&+J_{a}\sum_{\sigma _{1},\sigma _{2}}d_{x^{2}-y^{2},\sigma _{1}}^{\dagger
}d_{xy,\sigma _{2}}^{\dagger }d_{x^{2}-y^{2},\sigma _{2}}d_{xy,\sigma _{1}}
\nonumber \\
&&+J_{b}\sum_{\sigma _{1},\sigma _{2}}(d_{3z^{2}-r^{2},\sigma _{1}}^{\dagger
}d_{zx,\sigma _{2}}^{\dagger }d_{3z^{2}-r^{2},\sigma _{2}}d_{zx,\sigma _{1}}
\nonumber \\
&&\text{ \ \ \ \ \ \ \ \ \ \ \ }+d_{3z^{2}-r^{2},\sigma _{1}}^{\dagger
}d_{yz,\sigma _{2}}^{\dagger }d_{3z^{2}-r^{2},\sigma _{2}}d_{yz,\sigma _{1}})
\nonumber \\
&&+J_{e}\sum_{\sigma _{1},\sigma _{2}}d_{3z^{2}-r^{2},\sigma _{1}}^{\dagger
}d_{xy,\sigma _{2}}^{\dagger }d_{3z^{2}-r^{2},\sigma _{2}}d_{xy,\sigma _{1}}
\nonumber \\
&&+J_{t}[d_{x^{2}-y^{2},\uparrow }^{\dagger }d_{x^{2}-y^{2},\downarrow
}^{\dagger }(d_{zx,\downarrow }d_{zx,\uparrow }+d_{yz,\downarrow
}d_{yz,\uparrow })+\text{H.c.}]  \nonumber \\
&&+J_{a}[d_{x^{2}-y^{2},\uparrow }^{\dagger }d_{x^{2}-y^{2},\downarrow
}^{\dagger }d_{xy,\downarrow }d_{xy,\uparrow }+\text{H.c.}]  \nonumber \\
&&+J_{b}[d_{3z^{2}-r^{2},\uparrow }^{\dagger }d_{3z^{2}-r^{2},\downarrow
}^{\dagger }(d_{zx,\downarrow }d_{zx,\uparrow }+d_{yz,\downarrow
}d_{yz,\uparrow })+\text{H.c.}]  \nonumber \\
&&+J_{e}\sum_{\sigma _{1},\sigma _{2}}[d_{3z^{2}-r^{2},\uparrow }^{\dagger
}d_{3z^{2}-r^{2},\downarrow }^{\dagger }d_{xy,\downarrow }d_{xy,\uparrow }+%
\text{H.c.}]  \nonumber \\
&&+2\lambda \sum_{\sigma _{1},\sigma _{2}}(n_{yz,\sigma _{1}}-n_{zx,\sigma
_{1}})(d_{3z^{2}-r^{2},\sigma _{2}}^{\dagger }d_{x^{2}-y^{2},\sigma _{2}}+%
\text{H.c.})  \nonumber \\
&&+\lambda \sum_{\sigma _{1},\sigma _{2}}(d_{3z^{2}-r^{2},\sigma
_{1}}^{\dagger }d_{zx,\sigma _{2}}^{\dagger }d_{x^{2}-y^{2},\sigma
_{2}}d_{zx,\sigma _{1}}  \nonumber \\
&&\text{ \ \ \ \ \ \ \ \ \ \ }-d_{3z^{2}-r^{2},\sigma _{1}}^{\dagger
}d_{yz,\sigma _{2}}^{\dagger }d_{x^{2}-y^{2},\sigma _{2}}d_{yz,\sigma _{1}}+%
\text{H.c.})  \nonumber \\
&&+\lambda \lbrack (d_{zx,\uparrow }^{\dagger }d_{zx,\downarrow }^{\dagger
}-d_{yz,\uparrow }^{\dagger }d_{yz,\downarrow }^{\dagger })  \nonumber \\
&&\times (d_{x^{2}-y^{2},\downarrow }d_{3z^{2}-r^{2},\uparrow
}-d_{x^{2}-y^{2},\uparrow }d_{3z^{2}-r^{2},\downarrow })+\text{H.c.}],
\label{h2}
\end{eqnarray}
where $U=F_{0}+4F_{2}+36F_{4}$, $J_{e}=4F_{2}+15F_{4}$, $%
J_{t}=3F_{2}+20F_{4} $, $J_{a}=35F_{4}$, $J_{b}=F_{2}+30F_{4}$, and $\lambda
=\sqrt{3}(F_{2}-5F_{4})$.

The largest energy in $H_{I}$ is the intraorbital repulsion $U$. Most terms
of $H_{I}$ (in particular all those of the pure $e_{g}$ part $H_{4}$ and the
pure $t_{2g}$ part $H_{0}$) involve two and only two orbital indices, like
the (four different) ferromagnetic exchange interactions $J_{i}$, of the form

\begin{eqnarray}
\sum_{\sigma _{1},\sigma _{2}}d_{\xi ,\sigma _{1}}^{\dagger } d_{\eta
,\sigma _{2}}^{\dagger }d_{\xi ,\sigma _{2}}d_{\eta ,\sigma _{1}} = -(2
\mathbf{S_{\xi} \cdot S_{\eta}} +1/2)  \label{exch}
\end{eqnarray}
where $\mathbf{S_{\xi}}=\sum_{\chi \chi^{\prime}}d_{\xi, \chi }^{\dagger }
\mathbf{\sigma}_{\chi \chi^{\prime}}d_{\xi, \chi^{\prime}}/2$ is the spin of
the orbital $\xi$. The interorbital repulsion between these orbitals is
related to the former by $U^{\prime}=U-2J_{i}$ due to the spherical symmetry
of $H_I$. For the same reason, $J_{i}^{\prime }=J_{i}$, where $J_{i}^{\prime
}$ is the energy for transfer of intraorbital pairs (like $d_{\beta ,\uparrow }^{\dagger
}d_{\beta ,\downarrow }^{\dagger }d_{\gamma ,\downarrow }d_{\gamma ,\uparrow
}$). The remaining terms, with prefactor proportional to $\lambda $, involve
more than two orbitals. 
%To our knowledge terms like these were not considered before.

The parameters $F_{2}$ and $F_{4}$ related to the $J_{i}$ and $\lambda $ are
expected to be very weakly screened in the solid as compared to the free
atom and have little variation among the 3d series. For example, a fit of
the lowest atomic energy levels (given in Ref \ \cite{moore}) of V$^{+3}$
and Ni give nearly the same values $F_{2}=1300$ cm$^{-1}=0.16$ eV, and $%
F_{4}=88$ cm$^{-1}=0.011$ eV within 2\%. This results in an exchange energy $%
J_{e}=0.81$ eV for $e_{g}$ electrons and $J_{e}=0.70$ eV for $t_{2g}$ ones.
This implies for example that for two $e_{g}$ holes in cubic symmetry (as in
Ni$^{+2})$, the triplet ground state is separated by the excited singlet by 
$2J_{e}\simeq 1.6$ eV if covalency can be neglected. Recent calculations in
3d metals suggest that the exchange interactions are reduced in 30\% in
comparison with the atomic values.\cite{miya} This reduction is also assumed
in Ce compounds.\cite{haule}

In contrast to $F_{2}$ and $F_{4}$, $F_{0}$ which determines the
intra-orbital repulsion $U$ is significantly screened in the solids and is
difficult to determine theoretically. For the cuprates $U\simeq 10$ eV has
been estimated by constrained-density-functional calculations \cite%
{cdf1,cdf2} and it decreases to the left of the periodic table
inside the 3d series, because the 3d orbitals are more extended as
a consequence of the smaller nuclear charge. The value of $U$ can
be extracted from optical experiments.

\subsection{The full $M$-O Hamiltonian}

Since usually there are only a few O holes present, it turns out to be more
convenient to write the Hamiltonian in terms of hole operators (which
annihilate electrons) acting on the vacuum state in which all
transition-metal ($M$) ions are in the d$^{10}$ configuration and the O (or
P, As) ions are in the p$^{6}$ one. The most important of the remaining
terms of the starting Hamiltonian are the Co-O ($t_{\delta }^{\eta \xi }$
below) and O-O hopping ($\tau _{kj}^{\eta \vartheta }$), parameterized as
usual, in terms of the Slater-Koster parameters.\cite{slat}. We include a
cubic crystal field splitting $\epsilon _{t_{2g}}-\epsilon _{e_{g}}=10Dq$ at
the metal sites. Extension to tetragonal, tetrahedral or hexagonal crystal
fields is straightforward, while other symmetries may require a change in
the chosen basis for the d orbitals.

The Hamiltonian takes the form

\begin{eqnarray}
&&H =\sum_{i,\alpha \in e_{g},\sigma }\epsilon _{e_{g}}d_{i\alpha \sigma
}^{\dagger }d_{i\alpha \sigma }+\sum_{i,\beta \in t_{2g},\sigma }\epsilon
_{t_{2g}}d_{i\beta \sigma }^{\dagger }d_{i\beta \sigma }  \nonumber \\
&&+\sum_{j\eta \sigma }\epsilon _{j}p_{j\eta \sigma }^{\dagger }p_{j\eta
\sigma } +\sum_{i\delta \eta \xi \sigma }t_{\delta }^{\eta \xi }(p_{i+\delta
,\eta \sigma }^{\dagger }d_{i\xi \sigma }+\mathrm{H.c.})  \nonumber \\
&&+\sum_{j\neq k,\eta \vartheta \sigma }\tau _{kj}^{\eta \vartheta }p_{k\eta
\sigma }^{\dagger }p_{j\vartheta \sigma } +\sum_{i}H_{I}^{i}  \label{hs}
\end{eqnarray}
Here $p_{j\eta \sigma }^{\dagger }$ creates a hole on the O 2p orbital $\eta
$ at site $j$ with spin $\sigma $. The operator $d_{i\xi \sigma }^{\dagger }$
has an analogous meaning for the $M$ d orbitals at site $i$. The
interactions at this site $H_{I}^{i}$ has the form of Eqs. (\ref{his}) to (%
\ref{h2}) with the site index $i$ added to the subscripts of the $d$
operators. The subscript $i+\delta $ in $p_{i+\delta ,\eta \sigma }^{\dagger
}$ labels the different O atoms in the immediate neighborhood of the $M$
atom at $i$ (their nearest neighbors in highly symmetric structures).

We have neglected here for simplicity the on-site Coulomb repulsion at O
sites $U_{p}$ and the $M$-O interatomic repulsion $U_{pd}$. Experience in
the cuprates indicates that the former is not very important,\cite{bel}
while it complicates the numerical treatment of the basic $M$O$_{n}$ cluster
(see below). $U_{pd}$ can be incorporated easily and is important in the
formation of excitons \cite{bel2,exc} and in charge-transfer instabilities
\cite{cti} which are beyond the scope of this work.

\section{Construction of the effective Hamiltonian}

\subsection{Diagonalization of one cell}

The solution of the basic $M$O$_{n}$ cluster containing a transition-metal
atom and the O atoms in its neighborhood is greatly simplified for $U_{p}=0$%
, since among the $3n$ O orbitals, only some linear (bonding) combinations
with the same symmetry as the d orbitals hybridize with them, while the
remaining (non-bonding) orbitals decouple. The presence of $U_{p}$ would
introduce scattering between the bonding orbitals and the non-bonding ones
and does not modify the essential physics,\cite{bel} particularly for low or
moderate number of O holes.

For a perfect $M$O$_{6}$ octahedra (point group $O_{h}$) with $M$ at the
origin of cooordinates and the O atoms at $\delta =\pm \mathbf{x}$, $\pm
\mathbf{y}$, $\pm \mathbf{z}$, there are 13 O 2p non-bonding orbitals and
the 5 bonding O 2p orbitals are (spin indices and site index $i$ are omitted
for simplicity)

\begin{eqnarray}
p_{x^{2}-y^{2}} &=&\frac{1}{2}(p_{\mathbf{x},x}-p_{\mathbf{-x},x}-p_{\mathbf{%
y},y}+p_{\mathbf{-y},y}),  \nonumber \\
p_{3z^{2}-r^{2}} &=&\frac{1}{2\sqrt{3}}(2p_{\mathbf{z},z}-2p_{\mathbf{-z},z}
\nonumber \\
&&-p_{\mathbf{x},x}+p_{\mathbf{-x},x}-p_{\mathbf{y},y}+p_{\mathbf{-y},y}),
\nonumber \\
p_{xy} &=&\frac{1}{2}(p_{\mathbf{x},y}-p_{\mathbf{-x},y}+p_{\mathbf{y},x}
-p_{\mathbf{-y},x}),  \nonumber \\
p_{yz} &=&\frac{1}{2}(p_{\mathbf{y},z}-p_{\mathbf{-y},z}+p_{\mathbf{z},y}
-p_{\mathbf{-z},y}),  \nonumber \\
p_{zx} &=&\frac{1}{2}(p_{\mathbf{z},x}-p_{\mathbf{-z},x}+p_{\mathbf{x},z}
-p_{\mathbf{-x},z})  \label{p}
\end{eqnarray}
Writing the hopping terms in this basis, and using the Slater-Koster
formulas,\cite{slat} the bonding part of the Hamiltonian (\ref{hs}) in the 
$M $O$_{6}$ cluster takes the form

\begin{eqnarray}
H_{b} &=&\sum_{\alpha \in e_{g},\sigma }\{\epsilon _{e_{g}}d_{\alpha \sigma
}^{\dagger }d_{\alpha \sigma }+[\epsilon _{\text{O}}-2t_{p}]p_{\alpha \sigma
}^{\dagger }p_{\alpha \sigma }  \nonumber \\
&&-\sqrt{3}(pd\sigma )(d_{\alpha \sigma }^{\dagger }p_{\alpha \sigma }+\text{%
H.c.})\}  \nonumber \\
&&+\sum_{\beta \in t_{2g},\sigma }\{\epsilon _{t_{2g}}d_{\beta \sigma
}^{\dagger }d_{\beta \sigma }+[\epsilon _{\text{O}}+2t_{p}]p_{\beta \sigma
}^{\dagger }p_{\beta \sigma }  \nonumber \\
&&+2(pd\pi )(d_{\beta \sigma }^{\dagger }p_{\beta \sigma }+\text{H.c.}%
)\}+H_{I},  \label{hb}
\end{eqnarray}
where in terms of Slater-Koster parameters $t_{p}=-[(pp\sigma )-(pp\pi )]/2$.

If the cluster is elongated along the $z$ direction or O atoms at $\pm
\mathbf{z}$ are missing, in general one should consider two different 2p
orbitals that hybridize with $d_{3z^{2}-r^{2}}$. The first can be chosen as $%
(-p_{\mathbf{x},x}+p_{\mathbf{-x},x}-p_{\mathbf{y},y}+p_{\mathbf{-y},y})/2$,
while the second is $(p_{\mathbf{z},z}-p_{\mathbf{-z},z})/\sqrt{2}$ for $%
D_{4h}$ symmetry with obvious changes in absence of one or both O atoms at $%
\pm \mathbf{z}$. Similarly two independent states that hybridize with $%
d_{yz} $ in $D_{4h}$ symmetry are $(p_{\mathbf{y},z}-p_{\mathbf{-y},z})/%
\sqrt{2}$ and $(p_{\mathbf{z},y}-p_{\mathbf{-z},y})/\sqrt{2}$ and the same
changing $y $ by $x$. The hopping terms that involve these (at most 8) O
orbitals can be easily constructed using Ref. \onlinecite{slat}.

In any case, the size of the Hilbert space of $H_{b}$ with at most $13\times
2$ spin-orbitals is small enough to allow us to obtain its low-energy
eigenstates numerically by the Lanczos method.\cite{gag}

\subsection{The cobaltates}

For the case of Na$_{x}$CoO$_{2}$, we have solved numerically a CoO$_{6}$
cluster using the Hamiltonian $H_{b}$ given by Eq. (\ref{hb}), mapped the
corresponding states into those of an isolated Co atom with the
corresponding charge to calculate effective on-site interactions, and
calculated the hopping between effective Co sites mediated by O. The
parameters of $H_{b}$ were taken from fits to optical experiments 
\cite{kroll} as described below. We have neglected the trigonal distortion and
assumed $O_{h}$ symmetry in the cluster. As shown above, this assumption
reduces the size of the relevant Hilbert space and in addition simplifies
significantly the mapping between states of $H$ and $H_{\mathrm{eff}}$,
because as we shall see, the symmetry identifies unambiguously the
correspondence between states of both Hamiltonians. The effective trigonal
crystal-field splitting $\Delta =3D\sim 0.3$ eV was given by
quantum-chemistry configuration-interaction calculations.\cite{ll}
As we discuss in more detail below, this is one order of magnitude smaller
than the effective cubic crystal-field splitting and does not affect
our main findings.

Since Co$^{2+}$ is in a 3d$^{7}$ configuration (3 holes in the d shell), the
states of the CoO$_{6}$ cluster with $n+1$ holes are represented by a Co$%
^{n+}$ ion in $H_{\mathrm{eff}}$. The most relevant values of $n$ for Na$_{x}
$CoO$_{2}$, are 4 (the formal valence of Co for $x=0$) and 3 
(formal valence for the fully doped compound), but it is
important to consider also $n=5$ to calculate the effective interactions, as
we shall show. As the vacuum at one site for the effective Hamiltonian with
only Co sites, it is convenient to chose the Co$^{3+}$ 3d$^{6}$
configuration, occupied with the four $e_{g}$ holes. Thus Co$^{4+}$ has one $%
t_{2g}$ hole and Co$^{5+}$ has two $t_{2g}$ holes. Therefore, one expects
that the interacting part of $H_{\mathrm{eff}}$ at each site has the same
form as $H_{0}$ [Eq.(\ref{h0})] but now $U_{\mathrm{eff}}^{\prime}\neq U_{%
\mathrm{eff}}-2J_{\mathrm{eff}}$, and $J_{\mathrm{eff}}^{\prime }\neq J_{%
\mathrm{eff}}$ because the cubic crystal field $\epsilon _{t_{2g}}-\epsilon
_{e_{g}}=10Dq$ reduces the symmetry from that of the full rotational group
to $O_{h}$:

\begin{eqnarray}
H_{I}^{\mathrm{eff}} &=&U_{\mathrm{eff}}\sum_{\beta }\tilde{n}_{\beta
\uparrow }\tilde{n}_{\beta \downarrow }  \nonumber \\
&&+\frac{1}{2}\sum_{\gamma \neq \beta ,\sigma \sigma ^{\prime }}(U_{\mathrm{%
eff}}^{\prime }\tilde{n}_{\gamma \sigma }\tilde{n}_{\beta \sigma ^{\prime
}}+J_{\mathrm{eff}}\tilde{d}_{\gamma \sigma }^{\dagger }\tilde{d}_{\beta
\sigma ^{\prime }}^{\dagger }\tilde{d}_{\gamma \sigma ^{\prime }}\tilde{d}%
_{\beta \sigma })  \nonumber \\
&&+J_{\mathrm{eff}}^{\prime }\sum_{\gamma \neq \beta }\tilde{d}_{\gamma
\uparrow }^{\dagger }\tilde{d}_{\gamma \downarrow }^{\dagger }\tilde{d}%
_{\beta \downarrow }\tilde{d}_{\beta \uparrow },  \label{hie}
\end{eqnarray}%
This is the effective Hamiltonian at one site, except for an unimportant
constant $C$ and chemical potential term -$\mu N$, with $N=\sum_{\beta
\sigma }\tilde{n}_{\beta \sigma }$. The tilde above the operators is to
remind us that the effective operators $\tilde{d}_{\beta \sigma }^{\dagger }$
entering Eq. (\ref{hie}) are different from those entering the starting
Hamiltonian $H$, as discussed in more detail below. The resulting
eigenstates and energies of $H_{\mathrm{eff}}$ at one site, are listed in
the first three columns of Table I. It turns out that the low-energy
eigenstates of the CoO$_{6}$ cluster with 4, 5 and 6 holes are well
represented by the corresponding eigenstates of $H_{I}^{\mathrm{eff}}$. The
fourth column of Table I corresponds to the lowest energy levels of the
cluster Hamiltonian $H_{b}$ in each symmetry sector, using parameters
determined previously by us \cite{kroll} from a fit of polarized x-ray
absorption spectra of Na$_{x}$CoO$_{2}$ to the results of a CoO$_{6}$
cluster with 4 and 5 and holes including the core hole. In comparison with
the present calculations, the previous ones were simplified 
neglecting the exchange, transfer of intraorbital pairs, and terms proportional to $\lambda$ of 
$H_{1}$ and $H_{2}$, because their magnitude is smaller than the energy of the
crystal-field excitations related with these terms. The effect of
hybridization increases the splitting between $t_{2g}$ and $e_{g}$
orbitals to more than 3 eV $\gg F_2, F_4$.\cite{kroll} The effect
of the neglected terms is discussed in the next subsection.
The parameters of 
$H_{b}$ in eV are

\begin{eqnarray}
F_{0} &=&3.5\text{, }F_{2}=0.2\text{, }F_{4}=0.006\text{, }  \nonumber \\
(pd\pi ) &=&\frac{-\sqrt{3}}{4}(pd\sigma )=1\text{, }t_{p}=0.5,  \nonumber \\
\epsilon _{\text{O}} &=&13\text{, }\epsilon _{t_{2g}}=1.2\text{, }\epsilon
_{e_{g}}=0.  \label{par}
\end{eqnarray}
The choice of the origin of on-site energies $\epsilon _{e_{g}}=0$ is
arbitrary. The resulting value of $U=4.516$ eV and the charge transfer
\cite{note} energy $\Delta _{\text{CT}}=2.9$ eV are similar to those derived from
other x-ray absorption experiments.\cite{wu}

The most relevant eigenstates of the cluster for the physics of
the cobaltates are those corresponding to the first two rows of
Table I, corresponding to formal Co$^{3+}$ and Co$^{4+}$
respectively. These states are separated by at least 2 eV from
other eigenstates that cannot be represented by
$H_{\mathrm{eff}}$, like those containing non-bonding O orbitals
or states with intermediate spin.\cite{notes} This energy
difference is one order of magnitude larger than the corresponding
hopping amplitude, assuring the validity of $H^{\mathrm{eff}}$ as
representative of the low-energy physics of $H$. We note that
while in Co compounds, intermediate or high spin states are usual,
experiments show that the low spin states of Co are present in
Na$_x$CoO$_2$,\cite{wu,moto,lang,kro2} in agreement with our
results. This is due to the fact mentioned above, that in this system
the splitting between $t_{2g}$ and $e_{g}$ orbitals (due to $10Dq$ 
and Co-O hopping) is more than 3 eV, considerably larger than
the exchange energies.

From the six energies listed in the fourth column of Table I and their
corresponding expression for $H_{I}^{\mathrm{eff}}$ listed in the third
column, we have determined the four parameters of Eq. (\ref{hie}) and the
irrelevant constant $C$ and shift in chemical potential $\mu _{s}$. The
result in eV is

\begin{equation}
U_{\mathrm{eff}}=1.865\text{, }U_{\mathrm{eff}}^{\prime }=1.272\text{, }J_{%
\mathrm{eff}}=0.354\text{, }J_{\mathrm{eff}}^{\prime }=0.171  \label{pareff}
\end{equation}

\begin{widetext}

\begin{table}[t]
\caption{\label{tab:states} Eigenstates and energies of one site of
$H_{\rm eff}$, corresponding energies in the Co$O_6$ cluster and distribution
of holes in the latter. For degenerate representations only one state is shown. }
\begin{ruledtabular}
\begin{tabular}{llllllll}
Symmetry & eigenstate & $E-C+ \mu _{s} N$ & $E$ (eV) & d $e_g $& d $t_{2g}$ & p $e_g $& p $t_{2g}$ \medskip \\ \hline
%\vspace*{-10pt}
\\
$A^0_{1g}$ & $|0\rangle $ & 0 & 8.141 & 3.03 & 0.02 & 0.94 & 0 \medskip \\
$T^2_{2g}$ & $\tilde{d}_{xy \uparrow}^{\dagger }|0\rangle $ & 0 & 17.271 & 2.67 & 0.91 & 1.29 & 0.14 \medskip \\
$A^0_{1g} $ & $\frac{1}{\sqrt{3}}\sum_{\beta}\tilde{d}_{\beta \uparrow}^{\dagger }\tilde{d}_{\beta \downarrow }^{\dagger }
|0\rangle $ & $U_{\rm eff}+2J^{\prime }_{\rm eff}$ & 28.609 & 2.41 & 1.46 & 1.48 & 0.66 \medskip \\
$E^0 $ & $\frac{1}{\sqrt{2}}(\tilde{d}_{zx \uparrow}^{\dagger
}\tilde{d}_{zx\downarrow  }^{\dagger }-\tilde{d}_{yz\uparrow}^{\dagger }\tilde{d}_{yz\downarrow
}^{\dagger })|0\rangle $ & $U_{\rm eff}-J^{\prime }_{\rm eff}$ & 28.096 & 2.40 & 1.51 & 1.57 & 0.52 \medskip \\
$T^0_{2g}$ & $\frac{1}{\sqrt{2}}(\tilde{d}_{yz\uparrow }^{\dagger}\tilde{d}_{zx\downarrow }^{\dagger }
-\tilde{d}_{yz\downarrow }^{\dagger }\tilde{d}_{zx\uparrow}^{\dagger })|0\rangle $ & $U^{\prime }_{\rm eff}+J_{\rm eff}$
& 28.028 & 2.39 & 1.54 & 1.56 & 0.51 \medskip \\
$T^3_{1g} $ & $\tilde{d}_{xy\uparrow }^{\dagger }\tilde{d}_{yz\uparrow }^{\dagger
}|0\rangle $ & $U^{\prime }_{\rm eff}-J_{\rm eff}$ & 27.320 & 2.34 & 1.63 & 1.65 & 0.37 \medskip \\
\end{tabular}
\end{ruledtabular}
\end{table}

\end{widetext}

These parameters are smaller than those assumed in different calculations of
the electronic structure of Na$_{x}$CoO$_{2}$ near the Fermi surface which
include the effect of correlations.\cite{zhou,ishi,anto,mari,lieb,gut} For
example values of $U_{\mathrm{eff}}$ between 3 eV and $+\infty $ were used.
For our moderate values of the correlation energies, the hole pockets are
still predicted in theory in contrast to experiment if the small
first-principle value of the trigonal distortion $D$ is assumed.\cite%
{mari,lieb} Instead, using the value of $D$ given by quantum-chemistry
configuration-interaction calculations,\cite{ll} these pockets are absent
and the electronic dispersion near the Fermi energy agrees with experiment.%
\cite{dmft}

The reduction in the effective value of $U_{\mathrm{eff}}$ compared to the
3d value $U\sim 4.5$ eV is due to screening effects of the full model that
are ``hidden" in $H_{\mathrm{eff}}$. Actually, the operators $\tilde{d}%
_{\beta \sigma }^{\dagger }$ of $H_{\mathrm{eff}}$ do not correspond to pure
3d holes, but are complicated creation operators which involve 2p states.
The detailed expression of these effective operator in terms of those of $H$
is beyond the scope of the present work. Examples of the construction of
effective operators are given in Refs. \cite{opti,erol,spife}. In this
context, it is interesting to note that charge and current operators are
expressed as pure spin operators when Heisenberg-like models are used as
effective Hamiltonians for Hubbard-like models.\cite{ihm,cur} In any case,
it is easy to understand that if the effective 3d operators 
$\tilde{d}_{\beta \sigma }^{\dagger }$ have an important component of O 2p orbitals
distributed in the cluster, the effective repulsion between electrons
occupying these states is smaller than that between the corresponding
operators localized on the same Co ion.

In addition to covalency, another important point related with it and
noticed before by Marianetti \textit{et al.},\cite{mari2} is that the
addition of a new effective hole causes a redistribution of the remaining
charge. The distribution of charge among the different orbitals in the
cluster for the different states is listed in the last four columns of Table
I. The vacuum $|0\rangle $ of $H_{\mathrm{eff}}$ at one site, represents the
CoO$_{6}$ cluster with 4 holes. Near 3 of them occupy the 3d $e_{g}$ states
and the remaining one occupies mainly linear combinations of 2p O orbitals
with $e_{g}$ symmetry [corresponding to the first two lines of Eq. (\ref{p}%
)]. When an effective 3d $t_{2g}$ hole is added, nearly 0.35 3d $e_{g}$
holes are promoted to the O states with the same symmetry. A similar effect
is caused by the addition of the second effective 3d $t_{2g}$ hole (with
larger magnitude for the states of lowest energy). This is a combined effect
of the interorbital Coulomb repulsion, which increases the energy of the $%
e_{g}$ holes when $t_{2g}$ holes are added, and the strong hopping between
3d $e_{g}$ holes and linear combinations of O holes with the same symmetry.

The degree of covalency of the $t_{2g}$ states is less than that of the $%
e_{g}$ ones, due to the fact that the 3d-2p hopping is a factor 2 larger for
the latter [see Eqs. (\ref{hb}) and (\ref{par})].

\subsection{The effect of $H_{1}$ and $H_{2}$}
\label{h1h2}

Since most of the terms involved in Eqs. (\ref{h1}) and (\ref{h2})
are cumbersome, one might ask what happens if one neglects them.
The first four terms of $H_{2}$ (the simplest ones), containing
the Coulomb repulsion between d $e_{g}$ and $t_{2g}$ electrons are
crucial to obtain the right distribution of particles and should
be retained. Keeping the prefactors and neglecting the rest of the
terms of $H_{1}$ and $H_{2}$, the cubic $O_{h}$ symmetry is lost.
%For the parameters of the cobaltates that we used, some states of
%the former triply degenerate irreducible representations are
%separated form their center of gravity by about 0.06 eV.
Therefore, to retain the $O_{h}$ symmetry, we have replaced these
Coulomb repulsions by the average of all of them.

All energies increase with respect to those of the full
Hamiltonian. In particular for the states corresponding to nominal
Co$^{4+}$ (second row of Table I), the energy increases by 0.92
eV. However, the distribution of holes is not dramatically
affected. There is a 10 \% reduction in the amount of d~$t_{2g}$
from 0.91 to 0.82 which distribute evenly among other orbitals.
Similar effects take place for nominal Co$^{5+}$. The resulting
effective repulsions are decreased considerably. Proceeding as
indicated above we obtain $U_{\mathrm{eff}}=1.47$ eV,
$U_{\mathrm{eff}}^{\prime }=0.84$ eV (35\% less than for the full
Hamiltonian), $J_{\mathrm{eff}}=0.39$ eV, and
$J_{\mathrm{eff}}^{\prime }=0.25$ eV.

Of course, the changes should be more dramatic when the effective cubic
splitting ($10Dq$ plus effects of Co-O hopping) is smaller (like for other Co compounds),
since for spherical symmetry all terms of exchange and transfer of intraorbital pairs 
are equally important.

\subsection{The effective hopping}

\begin{figure}[tbp]
%[rbp]
\includegraphics[width=7cm]{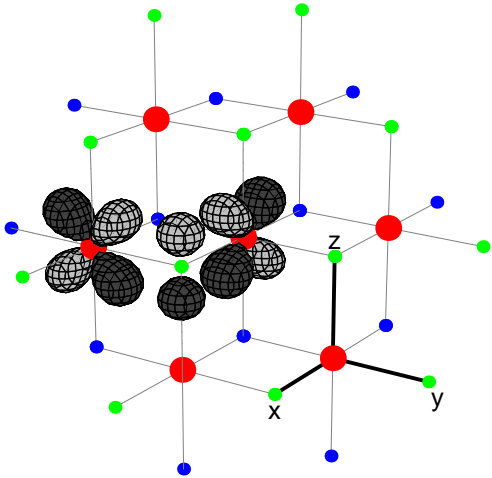}
\caption{(Color online) Scheme of the main orbitals involved in an effective
hopping from a d$_{yz}$ orbital at the left to a d$_{zx}$ orbital at the
right through an intermediate p$_{z}$ orbital.}
\label{hop}
\end{figure}

As shown first by Koshibae and Maekawa,\cite{koshi} the main hopping path
from one Co site to a neighboring Co site in Na$_{x}$CoO$_{2}$ is via an
intermediate O site. See Fig. \ref{hop}. The expression of this effective
hopping in second-order perturbation theory in the Co-O hopping is 
$t_{\mathrm{eff}}=(pd\pi )^{2}/\Delta _{\text{CT}}$. Using the parameters given
in Eq. (\ref{par}), $t_{\mathrm{eff}} \cong 0.34$ eV is obtained. Clearly,
this value is an overestimation, expected in covalent systems for which the
hopping term is not small enough compared to the charge-transfer energy. An
estimation based on the band width obtained from first principles gives $t_{%
\mathrm{eff}} \cong 0.1$ eV.\cite{kha}

It is known that the cell-perturbation method allows us to obtain more
accurate values of the effective hopping involving a calculation of only
first order in the perturbation, using the eigenstates of the 
cell.\cite{jef,brin,bel,bel2,hub,opti} Here we explain this procedure for 
Na$_{x}$CoO$_{2}$, using the eigenstates of the CoO$_{6}$ cluster with 5 holes
(represented by one occupied $t_{2g}$ orbital in $H_{\mathrm{eff}}$) and 4
holes (the vacuum in $H_{\mathrm{eff}}$).

To be specific, we start from a state in which the cluster $i$ (at
the left in Fig. \ref{hop}) is in its ground state
$|g_{i}(5,yz)\rangle $ with 5 holes and $yz$ symmetry and a
neighboring cluster $j$ (whose center corresponds to the rightmost
orbital represented in Fig. \ref{hop}) is in its non-degenerate
ground state $|g_{j}(4)\rangle $ with 4 holes (symmetry $A_{1g}$). 
After the hopping the final state is
$|g_{i}(4)g_{j}(5,zx)\rangle$. The initial and final states are
not orthogonal, but as we show below, this does not affect
significantly the resulting effective hopping. The spin is
conserved and the corresponding subscript is dropped for
simplicity. This
process is represented in $H_{\mathrm{eff}}$ by the hopping 
$\tilde{d}_{i,yz,\sigma }^{\dagger }\rightarrow \tilde{d}_{j,zx,\sigma }^{\dagger }$.
By symmetry, only the hopping terms of Eq. (\ref{hs}) \ which annihilate a
hole in the 2p$_{z}$ orbital of the common O atom 
($p_{i+\mathbf{y},z}=p_{j-\mathbf{x},z}$, middle orbital represented in Fig. \ref{hop}) contribute to 
$t_{\mathrm{eff}}$. Using Eq. (\ref{hs}) with Slater-Koster parameters, 
Eq.(\ref{p}) and symmetry one has (calling $t_{pd}=(pd\pi )$, $t_{p}=-[(pp\sigma
)-(pp\pi )]/2$)

\begin{eqnarray}
-t_{\mathrm{eff}} &=&\langle g_{i}(4)g_{j}(5,zx)|(t_{pd}d_{j,zx}^{\dagger
}+t_{p}(p_{j+\mathbf{z},x}^{\dagger }-p_{j-\mathbf{z},x}^{\dagger })
\nonumber \\
&&\times p_{j-\mathbf{x},z}|g_{i}(5,yz)g_{j}(4)\rangle   \nonumber \\
&=&\langle g_{j}(5,zx)|-t_{pd}d_{j,zx}^{\dagger }-t_{p}p_{j,zx}^{\dagger
}|g_{j}(4)\rangle   \nonumber \\
&&\times \langle g_{i}(4)|\frac{1}{2}p_{yz}|g_{i}(5,yz)\rangle   \label{teff}
\end{eqnarray}
In the equation above, the part of the Hamiltonian that leaves the initial state 
$|g_{i}(5,yz)g_{j}(4)\rangle$
unchanged is substracted [see Eq. (\ref{tort2}) below]. 
We have checked numerically that the perturbative expression 
$t_{\mathrm{eff}}=(pd\pi )^{2}/\Delta _{\text{CT}}$
is recovered for large  $\Delta _{\text{CT}}$.

To calculate the matrix elements, we have
used the simplified Hamiltonian, neglecting some terms of $H_{1}$ and $H_{2}$
as described above. With the parameters derived from optical experiments
[Eq. (\ref{par})], we obtain $t_{\mathrm{eff}}=0.101$ eV, which is in very
good agreement with the estimate of Ref. \onlinecite{kha}.

The final form of the non-interacting part of the effective Hamiltonian
contains the trigonal distortion $D\sim 0.1$ eV and a smaller direct hopping
between Co ions $t^{\prime }$.\cite{anto} These terms are beyond our
calculations based on the diagonalization of a CoO$_{6}$ cluster. The former
because we assumed $O_{h}$ symmetry and the latter because direct Co-Co
hopping cannot be included in the cluster. Nevertheless, the largest
energies in the problem are the interactions given by Eqs. (\ref{pareff})
within our approach.
Note that the value of the trigonal splitting $3D$ is an order of magnitude smaller 
than the effective cubic crystal field (near 3 eV including the effects of Co-O hybridization) 
and therefore does not sensibly affect the derivation of the remaining parameters 
of $H_{\mathrm{eff}}$, which were obtained assuming $O_h$ symmetry.

\subsection{The effect of the overlap}
\label{overlap}

While in $H_{\mathrm{eff}}$ one assumes a basis with orthogonal
states, the states of $H$ involved in the calculation of
$t_{\mathrm{eff}}$ above are non-orthogonal, due to the fact that
some p orbitals belong to different clusters. In the
cuprates,\cite{bel} and in double perovskites,\cite{petro} the
problem of non-orthogonality has been solved by a change of basis
to orthogonal Wannier functions centered at transition-metal
atoms. In the present case, in which five different 2p O Wannier
functions centered at each Co site should be considered, this
procedure cannot be handled analytically and becomes very awkward.
Fortunately, the larger number of orbitals involved has also the
consequence that the overlap between many-body states is
considerably reduced with respect to those of the one-particle
Wannier functions. This also takes place to a lesser extent in the
cuprates, in which the overlap between Zhang-Rice states is
S=-1/8, and
leads to a rapidly convergent expansion of $H_{\mathrm{eff}}$ in terms of S.%
\cite{sys}

Furthermore, a close examination of the structure of Na$_x$CoO$_2$
shows that in all
non-zero elements of the overlap matrix between bonding
combinations of O 2p orbitals [see Eq. (\ref{p})], at least one O
$t_{2g}$ orbital is involved. Since the occupation of the latter
is small in the relevant many-body eigenstates (see Table I), the
overlap between the latter is reduced. In particular we find

\begin{equation}
S=\langle g_{i}(4)g_{j}(5,zx)|g_{i}(5,yz)g_{j}(4)\rangle =\langle 1|2\rangle
=-0.0357,  \label{ove}
\end{equation}%
where for simplicity, we denoted as  $|1\rangle $ ($|2\rangle $), the
many-body state in which all clusters except $j$ ($i$) are in the ground
state for four holes and the remaining cluster is in the lowest lying state
for 5 holes and symmetry $zx$ ($yz$). To first order in the overlaps
$\langle l|m\rangle =S_{lm}$, $m\neq l$, orthonormal states can be obtained
using

\begin{equation}
|\tilde{l}\rangle =|l\rangle -\frac{1}{2}\sum\limits_{m,m\neq
l}S_{ml}|m\rangle .  \label{ort}
\end{equation}%
Then

\begin{equation}
t_{\mathrm{eff}}=\langle \tilde{1}|H|\tilde{2}\rangle =H_{12}-\frac{1}{2}%
S_{12}(H_{11}+H_{22}),  \label{tort}
\end{equation}%
where $H_{lm}=\langle l|H|m\rangle $. Since by symmetry $H_{11}=H_{22}$, the
above result can be written as

\begin{equation}
t_{\mathrm{eff}}=\langle 1|(H-H_{22})|2\rangle .  \label{tort2}
\end{equation}%
Except for the sign, the second member coincides with the second member of
Eq. (\ref{teff}). Therefore the result previously obtained is not modified
by terms linear in the overlap.

\section{Summary and discussion}

We have constructed the cell Hamiltonian $H_b$ that consists of a
transition-metal atom $M$ and its neighboring O atoms. This Hamiltonian is
the basis to construct a low-energy effective model $H_{\mathrm{eff}}$ in
which the O atoms are eliminated, using the cell-perturbation method.\cite%
{jef,brin,bel,bel2,hub,opti} The Hamiltonian has the same form if O is
replaced by other elements with p states near the Fermi energy, like P or As
in the Fe pnictides. An essential part of the cell Hamiltonian and the full
starting Hamiltonian $H$ is the interaction between electrons inside the d
shell $H_I$ when all 10 spin-orbitals are important. We have constructed $H_I
$ in a basis of orbitals appropriate for a cubic, hexagonal, tetragonal or
tetrahedral environment of the $M$ atoms. For other symmetries a change of
basis may be required.

While $H_I$ has a trivial form $U \sum_i n_{i \uparrow} n_{i \downarrow}$ in
the Hubbard model where only one orbital per site $i$ is relevant, the
interaction contains interorbital Coulomb repulsion, exchange and pair
hopping terms when only either $e_g$ [Eq.(\ref{h4})] or $t_{2g}$ [Eq.(\ref%
{h0})] are involved. These already makes the physics of transition-metal
compounds richer than that of the Hubbard model.\cite%
{dagh,oles,epl,oles2,fre,gus,vol,rodo}

If all orbitals should be retained, more complicated terms appear [Eqs. (\ref%
{h1}) and (\ref{h2})]. Our results suggest that because of the large $M$-O
hopping for $e_g$ holes in cubic or tetragonal symmetry, they should be
retained in general, even if the formal configuration of $M$ contains 6 d
electrons or less (corresponding to only $t_{2g}$ electrons in a ionic
picture).

In the case of the cobaltates Na$_{x}$CoO$_{2}$, where mainly 3d$^6$ and 3d$%
^5$ configurations play a role, our results as well as previous ones,\cite%
{mari2} show that addition or removal of electrons in the d shell causes a
strong charge redistribution between metal $e_g$ states and linear
combination of 2p O states with the same symmetry. This has important
consequences for the parameters of $H_{\mathrm{eff}}$. In particular, the
effective Coulomb repulsion $U_{\mathrm{eff}}$ is smaller than previously
assumed in calculations of $H_{\mathrm{eff}}$ which include the effect of
correlations.\cite{zhou,ishi,anto,mari,lieb,gut} As a consequence if the
hopping parameters and the trigonal distortion $D$ are obtained from a fit
of the bands obtained from first principles, six hole pockets appear along
the $\Gamma -K$ directions,\cite{mari,lieb} which are not detected in
photoemission experiments.\cite{arpes1,arpes2} Instead, using the values of
the interactions and effective hopping obtained as described in the previous
Section, and the larger value of $D$ given by quantum-chemistry
configuration-interaction calculations,\cite{ll} these pockets are absent
according to calculations which apply the dynamical mean-field theory (DMFT)
to $H_{\mathrm{eff}}$.\cite{dmft} The resulting electronic dispersion near
the Fermi energy agrees with experiment. Recent photoemission experiments in
misfit cobaltates show results similar as previous ones, with a significant
band reduction due to correlations.\cite{nico}

In general, the local-density approximation (LDA) underestimates gaps and
one-particle excitations energies. Thus one might suspect that it
underestimates the trigonal distortion energy $D$ in the cobaltates. The
above results suggest that taking the one-body parameters of the effective
model $H_{\mathrm{eff}}$ with metal sites only from a fit of the bands
obtained in LDA is not valid in general, or at least when the degree of
covalency is important. This is also the case of NiO, for which agreement
with experiment in LDA+DMFT calculations is only achieved once the O bands
are explicitly included in the model,\cite{vol} or when the O atoms have
been integrated out using low-energy reduction procedures similar as ours,
which take into account correlations from the beginning \cite{vol,oles}. The
research in the superconducting cuprates also supports the above statement.
In this case, the parameters of the effective one-band Hubbard or $t-J$
models are obtained accurately using systematic low-energy reduction
procedures \cite{jef,schu,brin,bel,bel2,sys,hub,opti} from a multiorbital
Cu-O model \cite{eme,var} with parameters obtained from
constrained-density-functional theory.\cite{cdf1,cdf2}

For the cobaltates, we have determined the parameters of the original
multiorbital model form a fit of polarized x-ray absorption spectra of 
Na$_{x}$CoO$_{2}$.\cite{kroll}

\section*{Acknowledgments}

We thank Antonin Bourgeois for useful discussions and a critical reading of the manuscript.
AAA is partially supported by CONICET, Argentina. This work was partially
supported by PIP 11220080101821 of CONICET, and PICT 2006/483 and PICT R1776
of the ANPCyT. TK has been financed by DFG grant KR 3611/1-1.


\begin{thebibliography}{99}
\bibitem{note} The charge transfer energy $\Delta _{\text{CT}}$ is defined
as the energy necessary to take a 3d hole of a Co$^{3+}$ ion and put it in a
2p orbital one of its neighboring O atoms in abscence of p-d hopping. See
Eq. (12) of Ref. \onlinecite{kroll}
\bibitem{notes} For example, the lowest eigenstate with spin 1 for four
holes in the cluster (corresponding to nominal Co$^{3+}$) has energy
$E=10.406$ eV. For nominal Co$^{4+}$ and S=3/2 we find $E=19.404$ eV.
\end{thebibliography}
\end{document}